\documentclass[aps,reprint,article,prx]{revtex4-1}
\usepackage{amsmath}
\usepackage{dcolumn}
\usepackage{graphicx}
\usepackage{float}
\usepackage{multirow}

\begin{document}
\title{Single particle tunneling spectroscopy and superconducting gaps in layered iron based superconductor KCa$_{2}$Fe$_{4}$As$_{4}$F$_{2}$}
\author{Wen Duan,$^1$ Kailun Chen,$^1$ Wenshan Hong,$^{2,3}$ Xiaoyu Chen,$^1$ Huan Yang,$^{1,*}$ Shiliang Li,$^{2,3,4}$ Huiqian Luo,$^{2,4}$ and Hai-Hu Wen$^{1,\dag}$}

\affiliation{$^{1}$National Laboratory of Solid State Microstructures and Department of Physics, Collaborative Innovation Center of Advanced Microstructures, Nanjing University, Nanjing 210093, China}

\affiliation{$^{2}$Beijing National Laboratory for Condensed Matter Physics, Institute of Physics, Chinese Academy of Sciences, Beijing 100190, China}

\affiliation{$^{3}$School of Physical Sciences, University of Chinese Academy of Sciences, Beijing 100190, China}

\affiliation{$^{4}$Songshan Lake Materials Laboratory, Dongguan, Guangdong 523808, China}

\begin{abstract}
We perform scanning tunneling microscopy/spectroscopy study on the layered iron based superconductor KCa$_2$Fe$_4$As$_4$F$_2$ with a critical temperature of about 33.5 K.
Two types of terminated surfaces are generally observed after cleaving the samples in vacuum. On one commonly obtained surface, we observe a full gap feature with energy gap values close to 4.6 meV. This type of spectrum shows a clean and uniform full gap in space, which indicates the absence of gap nodes in this superconductor. Quasiparticle interference patterns have also been measured, which show no scattering patterns between the hole and tiny electron pockets, but rather an intra-band scattering pattern is observed possibly due to the hole-like $\alpha$ pocket. The  Fermi energy of this band is only about $24\pm6$ meV as derived from the energy dispersion result. Meanwhile, impurity induced bound-state peaks can be observed at about $\pm2.2$ meV on some spectra, and the peak value seems to be independent to magnetic field. On the second type of surface which is rarely obtained, the fully gapped feature can still be observed on the tunneling spectra, although multiple gaps are obtained either from a single spectrum or separate ones, and the gap values determined from coherence peaks locate mainly in the range from 4 to 8 meV. Our results clearly indicate multiple and nodeless superconducting gap nature in this layered superconductor KCa$_2$Fe$_4$As$_4$F$_2$, and the superfluid is mainly contributed by the hole-like Fermi surfaces near $\Gamma$ point. This would inspire further consideration on the effect of the shallow and incipient bands near M point, and help to understand the pairing mechanism in this highly layered iron-based superconductor.
\end{abstract}

\maketitle
\section{Introduction}
Iron-based superconductors (FeSCs) are the second family of unconventional high-temperature superconductors. In most FeSCs, several Fe derivative $d$-bands cross the Fermi energy forming the electron- and hole-like pockets. Meanwhile, band structures and Fermi surfaces are quite different in various FeSCs, and they are also very sensitive to chemical doping or external pressure. The widely accepted $s^\pm$ pairing symmetry in some FeSCs is based on the nesting between hole pockets near $\Gamma$ point and electron pockets around M point with similar sizes based on the weak coupling scenario, but the gap symmetry and the gap structure can be different in other FeSCs because of different structures of Fermi surfaces \cite{Review}.

The newly found $A\mathrm{Ca_2Fe_4As_4F_2}$ ($A =$ K, Rb, Cs) is a representative compound of the layered FeSCs, and the critical temperature ($T_\mathrm{c}$) ranges from 28 to 33 K \cite{K12442,RbCs12442,Growthmethod}. The crystal structure of $\mathrm{KCa_2Fe_4As_4F_2}$ (K12442) is shown in Fig.~\ref{fig1}(a) as an example, one can see that in these materials, double $\mathrm{FeAs}$ layers are separated by insulating $\mathrm{Ca_2F_2}$ layers.  Such kind of layered structure results in a significant anisotropy of superconductivity and normal-state resistance \cite{Growthmethod,anisotropytransport,magneticmeasurement,anisotropicJc}. It is supposed that the 12442-type FeSCs have the intrinsic hole conduction with the doping level of 0.25 hole/Fe. Interestingly, it is easy to transform the primary carrier from $p$-type to $n$-type by Co or Ni doping, but $T_\mathrm{c}$ decreases with increase of the doped concentration of Co or Ni dopants \cite{Codoping,Nidoping}. Meanwhile, $T_\mathrm{c}$ can be slightly enhanced by applying a hydrostatic pressure \cite{pressureenhancement}. Transport measurements in $\mathrm{KCa_2Fe_4As_4F_2}$ (K12442) single crystals suggest that the in-plane upper critical field is dominated by the Pauli paramagnetic effect instead of the orbital effect \cite{StrongPaulieffect}. Theoretical calculation predicts that there are several hole and electron pockets in the K12442 \cite{Codoping,firstprinciple,Bandcaculation}. Based on a recent angle-resolved photoemission spectroscopy (ARPES) work conducted in K12442, three separate hole pockets $\alpha$, $\beta$, $\gamma$ are observed around the $\Gamma$ point, and one tiny electron pocket together with four incipient hole bands (which barely touch the Fermi energy) are observed around the M point \cite{ARPES}. Obviously, this topology of Fermi surface cannot satisfy the nesting condition because of very different sizes of hole and electron pockets. The nesting condition is satisfied by Co doping to K12442 with the doping level of about 0.1, but $T_\mathrm{c}$ decreases to about 25 K \cite{Codoping}. In this point of view, the superconductivity in 12442-type FeSCs may be different from other FeSCs. ARPES measurements in K12442 exhibit six nodeless gaps with gap values ranging from 2 meV to 8 meV for different Fermi pockets. The multiple and nodeless gap feature are also proved by different experimental methods \cite{HeatTran,optical}. However, some other works claim that there might be line nodes on the superconducting gap(s) in 12442-type FeSCs \cite{uSRCs,uSRK,specific heat}. The controversies of the existence of gap nodes in 12442 system require further investigations. Although the nesting condition is not satisfied in 12442-tpye FeSCs, the spin resonance peak is still observed around $Q=(0.5,0.5)$ \cite{NSRK,NSRCs} which corresponds to the scattering vector from the hole to electron pockets. Here, the $s^\pm$ pairing symmetry with the spin resonance can be explained in the strong coupling approach with the absence of the nesting condition \cite{DaiReview}. In addition, the spin resonance mode with a downward dispersion is observed in K12442, and this kind of dispersion is similar to the behavior in cuprates \cite{NSRK}.

In this paper, we report the experimental study on $\mathrm{KCa_2Fe_4As_4F_2}$ single crystals by the scanning tunneling microscopy/spectroscopy (STM/STS). Fully gapped feature is observed on almost all tunneling spectra. We also conduct the quasiparticle interference (QPI) measurements in the sample in order to obtain the information of Fermi pockets. Our results provide fruitful information to this multi-band superconductor.

\section{Experimental Methods}

\begin{figure}[htbp]
\centering
\includegraphics[width=8cm,height=8.5cm]{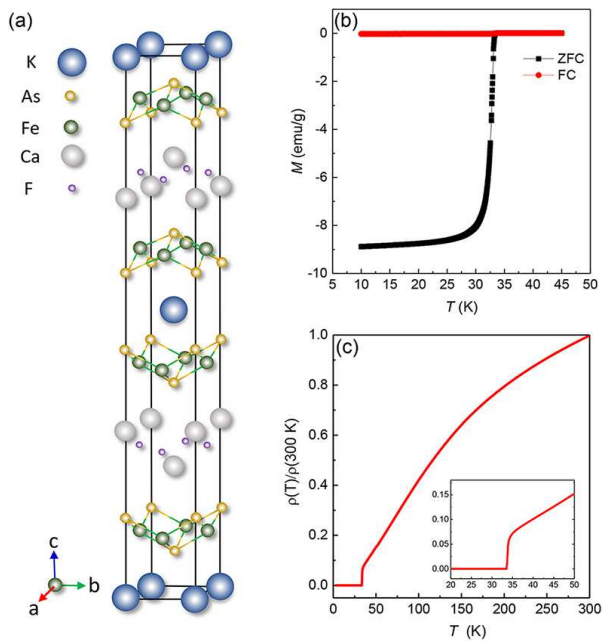}
\caption{(a) Crystal structure of $\mathrm{KCa_2Fe_4As_4F_2}$. (b) Temperature dependent magnetization measured in zero-field cooled (ZFC) and field-cooled (FC) processes under a magnetic field of 10 Oe. (c) Temperature dependence of normalized in-plane resistance measured at 0 T.
} \label{fig1}
\end{figure}

The KCa$_2$Fe$_4$As$_4$F$_2$ single crystals used in this work were grown by the self-flux method \cite{Growthmethod}. Temperature dependent magnetization and normalized resistance are shown in Figs.~\ref{fig1}(b) and \ref{fig1}(c), and both of them show fine superconducting transitions with critical temperature $T_\mathrm{c}$ of about 33.5 K determined from the zero-resistance. STM/STS measurements were carried out in a scanning tunneling microscope (USM-1300, Unisoku Co., Ltd.). The K12442 samples were cleaved at about 77 K in an ultrahigh vacuum with the base pressure of about $1\times10^{-10}$ Torr, and then they were transferred to the microscopy head which was kept at a low temperature. Electrochemically etched tungsten tips were used for STM/STS measurements after cleaning by the electron-beam heating. A typical lock-in technique was used in tunneling spectrum measurements with an ac modulation of 0.1 mV and the frequency of 931.773 Hz. Voltage offsets were carefully calibrated before STS measurements.

\section{Results}

\subsection{Topography and tunneling spectra}

\begin{figure}[H]
\centering
\includegraphics[width=8cm]{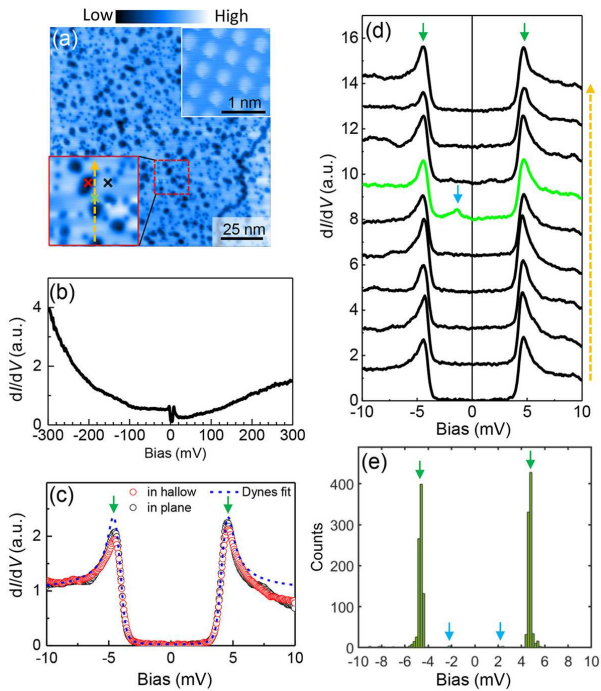}
\caption{(a) A typical topographic image taken on a surface of K12442 measured at $T = 1.7$ K with setpoint conditions of $V_\mathrm{set} = 20$ mV and $I_\mathrm{set} = 200$ pA. The inset in the upper-right corner shows the atomically resolved topography measured in another flat area ($V_\mathrm{set} = 10$ mV and $I_\mathrm{set} = 500$ pA). The inset in the lower-left corner shows the re-scanned image with a higher resolution of an area in (a) marked by the red dashed square ($V_\mathrm{set} = 20$ mV, $I_\mathrm{set} = 200$ pA). (b) A typical tunneling spectrum measured in a energy window far beyond the superconducting gap ($T = 1.7$ K, $V_\mathrm{set}$ = 300 mV, and $I_\mathrm{set}$ = 500 pA). (c) Two tunneling spectra measured at the marked positions by the red and black crosses in the lower-left inset in (a) ($T = 1.7$ K, $V_\mathrm{set}$ = 10 mV, and $I_\mathrm{set}$ = 200 pA). The spectrum in red color is measured at the center of the red cross in the hollow, while the spectrum in black color is measured at the center of black cross in the flat area. The blue dashed line shows the fitting curve by the Dynes model with a slightly anisotropic $s$-wave gap function. (d) A spatially resolved tunneling spectra measured along the yellow dashed line in the lower-left inset in (a) ($T = 0.4$ K, $V_\mathrm{set}$ = 20 mV, and $I_\mathrm{set}$ = 200 pA). The positions of the coherence peaks are marked by the green arrows. The spectrum with green color is measured at the center of the green cross shown in the lower-left inset in (a), which exhibits a peak at bias voltage of about -2 mV (marked by a blue arrow), and it may be the impurity induced state. (e) The statistics of peak energies in 845 tunneling spectra measured at randomly selected points in the area of (a) ($T = 1.7$ K, $V_\mathrm{set} = 10$ mV, and $I_\mathrm{set} = 200$ pA). The blue arrows indicate the existence of low-energy peaks at about $\pm2.2$ meV, and they may be induced by impurities.
} \label{fig2}
\end{figure}

Figure~\ref{fig2}(a) shows a typical topographic image measured on the surface of K12442 single crystal. Based on the lattice structure of K12442, there are layers of alkali-metal K atoms and those of alkaline-earth-metal Ca atoms. The cleavage may occur in these layers with the relatively weak bonding energy. After the cleavage, most probably, half K or Ca atoms remain in the surface layer of each separated part, which makes both surface unpolarized. This can get a proof from an atomically resolved topography shown in the upper-right inset in Fig.~\ref{fig2}(a) measured on a flat area far away from any defects. The topography shows a square lattice with the lattice constant of about 5.3 \AA\ which is approximately equal to $\sqrt2$ times of the K-K or Ca-Ca lattice constant ($a_0=3.87$ \AA). From the topographic image, one can see that there are many hollows with different sizes on the flat background. The depths of the hollows are from 100 to 300 pm, and these hollows can be clearly seen from the re-scanned image shown in the lower-left inset in Fig.~\ref{fig2}(a). Similar kinds of hollows have been observed in NaFe$_{1-x}$Co$_x$As \cite{NaFeCoAs}, LiFeAs \cite{LiFeAs}, and RbFe$_2$As$_2$ \cite{RbFe2As2} but with much lower densities, and they may be the assembled vacancies of alkali metal atoms on the reconstructed surface. In Fig.~\ref{fig2}(b), we show a typical tunneling spectrum on the surface measured in a wide energy window. The differential conductance is much larger in negative-bias side than that in positive-bias side, which is consistent with the asymmetric density of states from previous band calculation results \cite{firstprinciple,Bandcaculation}.

Figure~\ref{fig2}(c) shows two tunneling spectra measured at two marked positions in the lower-left inset in Fig.~\ref{fig2}(a), i.e., one is measured at a position in a hollow, and the other is measured at a position on the flat area far away from the hollows. One can see that the two spectra show almost the same feature, which suggests that hollows have very little influence on the superconductivity. A slight suppression of the intensity of coherence peaks can be observed on the spectrum measured in the hollow compared to that measured in the flat area. Both of the two spectra show a full gap feature with a pair of coherence peaks locating at energies of about $\pm$4.6 meV. Then we carry out the Dynes model \cite{Dynes} with an $s$-wave gap to fit the spectrum measured in the flat area. The best fitting result is shown as the dashed curve in Fig.~\ref{fig2}(c), and it requires a slightly anisotropic $s$-wave gap for the best fitting. The obtained gap function reads $\Delta(\theta) = 4.6(0.93+0.07\cos2\theta)$ meV, and the scattering rate $\Gamma = 0.1$ meV. Here the gap maximum $\Delta_\mathrm{max}$ is close to the energy value of coherence peaks, and it is also similar to gap values of hole pockets of $\alpha$ and $\beta_1$ or the electron pocket of $\delta$ from the ARPES measurements \cite{ARPES}. We also measured a set of tunneling spectra along a dashed line in the lower-left inset in Fig.~\ref{fig2}(a), and the spectra are shown in Fig.~\ref{fig2}(d). All the spectra are homogeneous except for a slight change of the coherence peak energy. On the spectrum in green color shown in Fig.~\ref{fig2}(d), one can see that there is a small peak at about 2 mV marked by a blue arrow. It should be noted that this spectrum is not measured in the hollow, and there is no any unique feature on the topography. The peak is likely to be the bound state induce by an impurity underneath the surface, which will be discussed below in Subsection~\ref{impurity}. We then conduct the tunneling spectrum measurement all over the area in Fig.~\ref{fig2}(a), and do the statistics to peak energies on 845 measured spectra. The result is shown in Fig.~\ref{fig2}(e). The coherence peaks are mostly located from 4.4 to 5.4 meV from the statistics. About 2\% of spectra have low-energy peaks within the energy range of $\pm(2.2\pm0.4)$ mV, they can appear either at the hollow positions or in the flat area.

\subsection{Results of quasiparticle interference}

\begin{figure}[htbp]
\centering
\includegraphics[width=8cm]{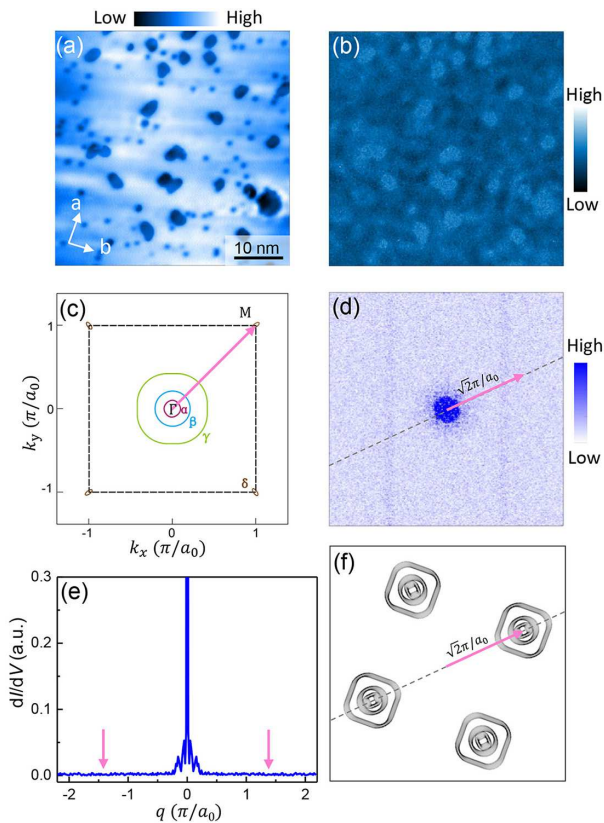}
\caption{(a) Topographic image ($V_\mathrm{set}$ = 30 mV, $I_\mathrm{set}$ = 200 pA) and (b) the corresponding normal-state QPI mapping at $E = 10$ meV ($V_\mathrm{set} = 30$ mV, $I_\mathrm{set} = 200$ pA) measured in the same area. (c) Schematic plot of Fermi surfaces derived from a previous ARPES work \cite{ARPES}. (d) The Fourier transformed pattern of the QPI mapping in (b). (e) Line profile plot of the intensity of the FT-QPI pattern along the black dashed line in (d), and the arrows indicate the position of scattering vectors connecting $\Gamma$ and M points. (f) The simulated scattering patterns between the hole and electron pockets plotted as grey patterns with center of $\sqrt{2}\pi/a_0$ from the center of the FT-QPI pattern. Here the grey patterns are the selected patterns in the self-correlated image of (c). And the arrow represents the scattering vector connecting $\Gamma$ and M points. Comparing (d) and (f), one can see that the scattering between hole and electron pockets has not been detected from our experimental data. All measurements are carried out at 0.7 K.
} \label{fig3}
\end{figure}

\begin{figure*}[htbp]
\centering
\includegraphics[width=18cm,height=13cm]{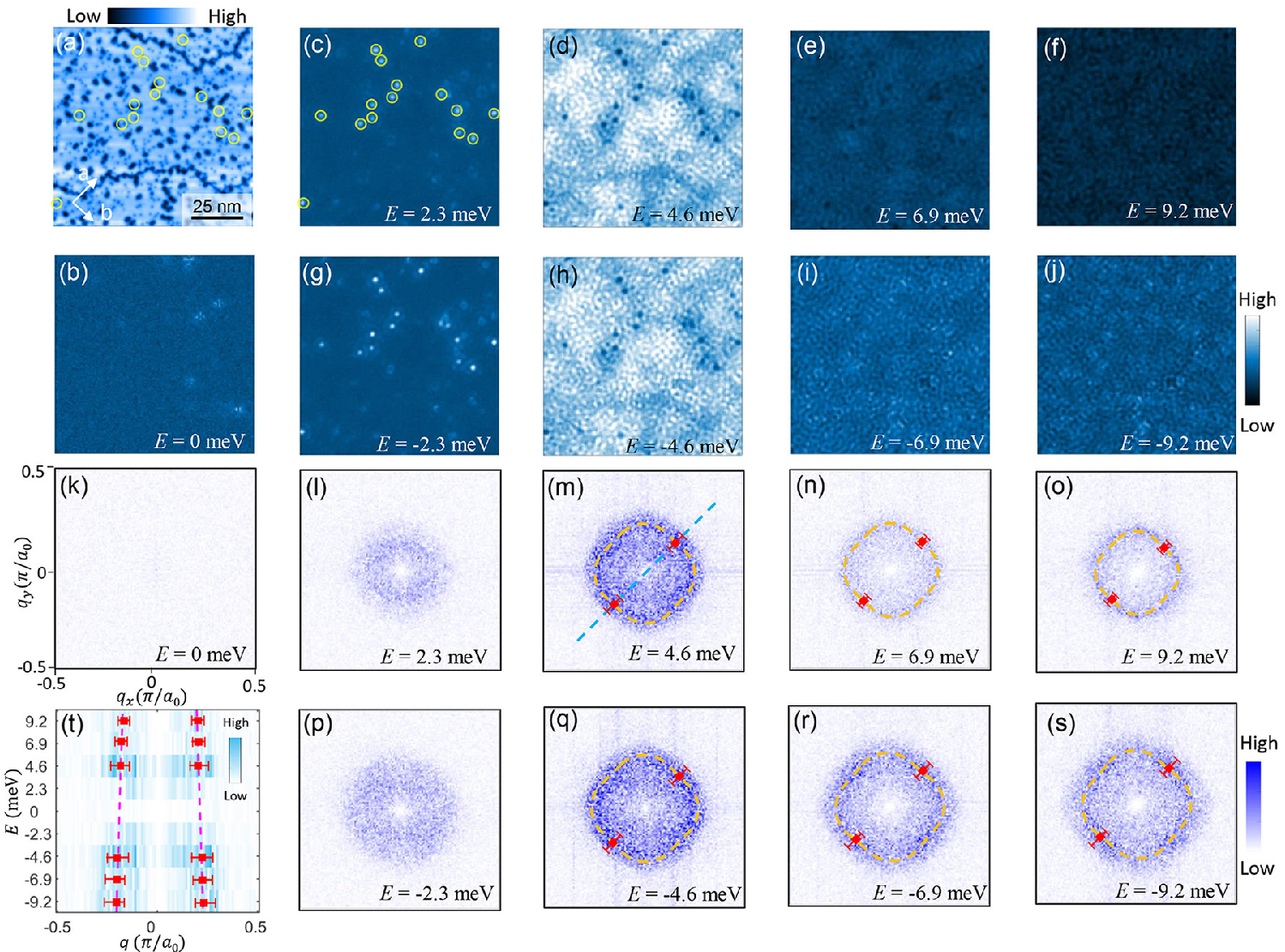}
\caption{(a) A large-scale topographic image measured on the surface of K12442 ($V_\mathrm{set}$ = 10 mV, $I_\mathrm{set}$ = 200 pA). (b)-(j) The QPI mapping measured in (a) at different energies with $B = 0$ T and $T= 1.1$ K ($V_\mathrm{set}$ = 10 mV, $I_\mathrm{set}$ = 200 pA). Circles in (a) and (c) mark positions of impurities with the bound state energy at about $\pm2.3$ mV. The locations of these impurities seem to be irrelevant to the hollows on the top surface. (k)-(s) Corresponding FT-QPI patterns derived from Fourier transformation to the QPI mappings (b)-(j), respectively. A 2D-Gaussian-function background is subtracted in the center of all the FT-QPI patterns, and the full width at half maximum is about $0.18\pi/a_0$ which is a very small value. (t) The energy dispersion of intensity in FT-QPI patterns along the diagonal direction marked by a blue dashed line in (m). The energy dispersion feature seems to be hole-like. The dashed curve is a guide line of a parabola, and we can obtain the energy $E_b\approx24\pm6$ meV for the top of the band. Red full squares represent sizes of the FT-QPI patterns marked as orange dashed circles in (m)-(o) and (q)-(s). The error bars highlighted in (t) reflect roughly the estimated width of the circle-like outline of the FT-QPI pattern.
} \label{fig4}
\end{figure*}

QPI measurements and the related analysis are very useful because they can tell the information of the Fermi surface \cite{Fe-STM-Hoffman}, the gap anisotropy \cite{LiFeAsani,FeSe11111ani}, as well as the gap signs \cite{HanaguriScience,LiFeAss+-,FeSes+-,FeSe11111s+-,Bi2212d} in a superconductor. We also measure the differential conductance mapping and show a QPI mapping at $E = 10$ meV in Fig.~\ref{fig3}(b). Although the hollows in the topography do not affect the superconductivity too much, standing waves can be clearly seen surrounding these hollows. When we do the Fourier transformation to the QPI mapping, we can obtain the Fourier-transformed (FT-) QPI pattern and show it in Fig.~\ref{fig3}(d). In K12442, a previous ARPES work \cite{ARPES} observe three hole pockets ($\alpha$, $\beta$, and $\gamma$) around the $\Gamma$ point and one tiny electron $\delta$ pocket around M point. We plot a sketch map of Fermi surfaces in Fig.~\ref{fig3}(c). Here the intra band scattering should locate around the center point of the FT-QPI pattern shown in Fig.~\ref{fig3}(d), and the simulated scattering results between the hole and the electron pockets are plotted as the four grey patterns in Fig.~\ref{fig3}(f). However, these scattering patterns have not been observed in the experimental data shown in (d). It is more clear in the line-cut intensity of the differential conductance shown in Fig.~\ref{fig3}(e). It is almost featureless near the points of $q=\pm\sqrt{2}\pi/a_0$ which connects $\Gamma$ and M points in momentum space. The absence of the characteristic scattering patterns between the hole and electron pockets may be explained by following two possible reasons. (a) A low density of states at the Fermi energy for the small electron $\delta$ pockets; (b) The unsensitive tunneling matrix element effect. A detailed discussion is given in the Section~\ref{discussion}.

Since we have not observed the scattering between the hole and electron pockets, we try to get some information of the intra-band scattering from the QPI data in a large area. Results of QPI mappings and corresponding FT-QPI patterns are shown in Fig.~\ref{fig4}. At zero energy, almost no clear features can be observed in QPI pattern in Fig.~\ref{fig4}(b), which is consistent with the full gap feature from tunneling spectrum measurements. At $E=\pm2.3$ meV, one can see obviously in Fig.~\ref{fig4}(c) that there are about 15 clear spots induced by the impurity bound state at this energy. When we go back to the topographic image in Fig.~\ref{fig4}(a), we can conclude that there is no evident correlation between impurity locations and the hollows on the top surface. The impurity may be beneath the top surface, e.g., in FeAs layer. The most frequent superconducting gap value or the coherence peak energy is about 4.6 meV from tunneling spectrum measurements on this surface, and it is comprehensible that the intensity of FT-QPI patterns are very strong at $E=\pm4.6$ meV. Although we cannot distinguish the detailed scattering features, it is clear that there are fourfold diamond-like patterns around the center in FT-QPI patterns. The traces of these outlines arising from the FT-QPI patterns are marked by yellow dashed lines in Figs.~\ref{fig4}(m)-(o) and \ref{fig4}(q)-(s). The size of such fourfold diamond-like pattern shrink with increase of the energy, which can be clearly seen in the energy dispersion plot [Fig.~\ref{fig4}(t)]. This result suggests that the scattering may be due to the intra-band scattering between the hole pockets near $\Gamma$. When we try to fit the dispersion data with a parabola, an energy value of the band top can be estimated to be about $24\pm6$ meV, and the diameter of the relevant Fermi pocket is about $0.18\pi/a_0$. These values are close to the parameters of the hole-like $\alpha$ pocket as determined from ARPES measurements \cite{ARPES}.

\subsection{Impurity bound states}\label{impurity}

\begin{figure}[htbp]
\centering
\includegraphics[width=8cm,height=9cm]{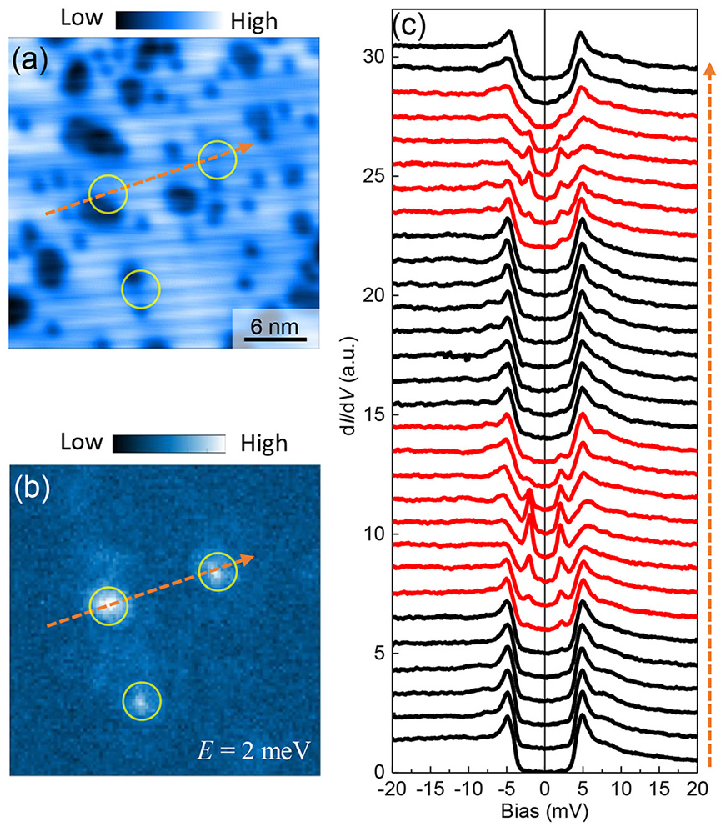}
\caption{(a) Topographic image ($V_\mathrm{set}$ = 20 mV, $I_\mathrm{set}$ = 200 pA) and (b) QPI mapping at $E = 2$ meV ($V_\mathrm{set}$ = 20 mV, $I_\mathrm{set}$ = 200 pA) measured in the same area. One can see three bright spots in (b). (c) A set of tunneling spectra ($V_\mathrm{set}$ = 20 mV, $I_\mathrm{set}$ = 200 pA) measured along the arrowed line in (a) or (b) crossing centers of these two spots. The tunneling spectra in red are the ones measured in the area of the bright spot. All measurements are carried out at 1.7 K.
} \label{fig5}
\end{figure}

\begin{figure}[htbp]
\centering
\includegraphics[width=8cm]{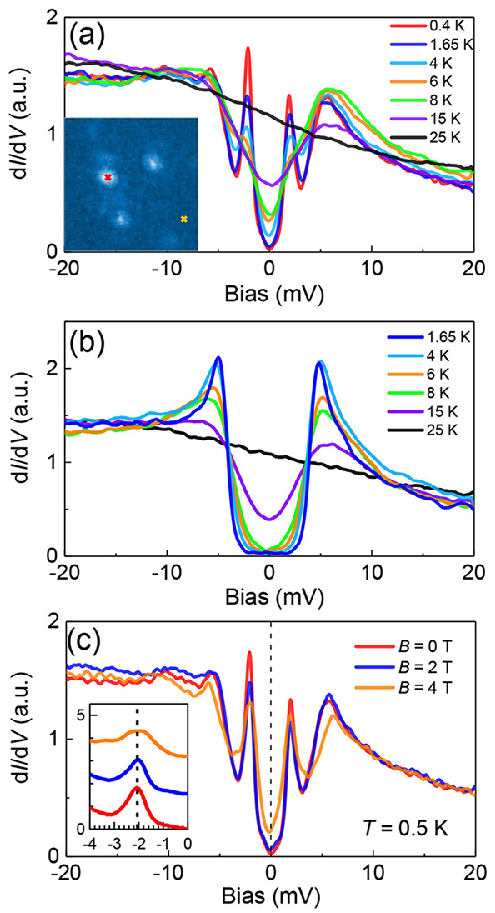}
\caption{Temperature dependent evolution of the tunneling spectra measured (a) at the bright-spot center and (b) far away from any impurities. Specifically, (a) and (b) are measured at centers of the red and yellow crosses marked in the inset of (a), respectively. (c) Magnetic field evolution of the tunneling spectra measured at the impurity center. The inset shows the enlarged view of the bound state peak at about $-2$ meV. Setpoint conditions: $V_\mathrm{set}$ = 20 mV, $I_\mathrm{set}$ = 200 pA.
} \label{fig6}
\end{figure}

In order to investigate low-energy peaks at about $\pm2.2$ meV, we measured spatial evolution of tunneling spectra [Fig.~\ref{fig5}(c)] along an arrowed line across two bright spots in QPI mapping [Fig.~\ref{fig5}(b)]. One can see in Fig.~\ref{fig5}(c) that the tunneling spectra have the similar feature of peaks when across these two spots. Such peaks can be even sharper at 0.4 K, and one example can be seen in Fig.~\ref{fig6}(a). The spectrum feature near zero bias can even be affected by these low-energy peaks. The extremely sharp peaks may exclude the possibility of a smaller superconducting gap, because we can not fit the experimental data well by using the Dynes model with two superconducting gaps. In addition, the peaks at about $\pm2$ meV disappear when the temperature is 8 K, but the coherence peaks at about $\pm5$ meV can exist at temperature above 15 K. Following the discussions mentioned above, we argue that low-energy peaks are impurity induced bound states although these impurities locate underneath. In Fig.~\ref{fig6}(c), we show tunneling spectra measured at the center of a bright spot and under different magnetic fields. The amplitude of the impurity induced peak lowers down with increase of the magnetic field. The inset of Fig.~\ref{fig6}(c) shows the enlarged view of the impurity induced peak in negative-energy side, and the peak energy is almost unchanged under the magnetic field of 4 T. From a previous report, the field-induced peak-shift slope is about 0.06 meV/T for the impurity bound state induced by a magnetic Fe-vacancy impurity with the Land\'{e} factor $g=2$ \cite{XueFevacancy}. Based on this slope, we can calculate the energy shift is about 0.24 meV when the field changes from 0 to 4 T. However, the energy shift is negligible for the impurity bound state in the K12442 sample, we can argue that the impurity may be non-magnetic or weak magnetic. In addition, the peak feature is similar to the bound state peak of the As vacancy \cite{AsVancancy}, so impurities are likely to be non-magnetic As vacancies in the FeAs layer underneath.

\subsection{Tunneling spectra measured on another kind of terminated surface}

\begin{figure}[htbp]
\centering
\includegraphics[width=8cm]{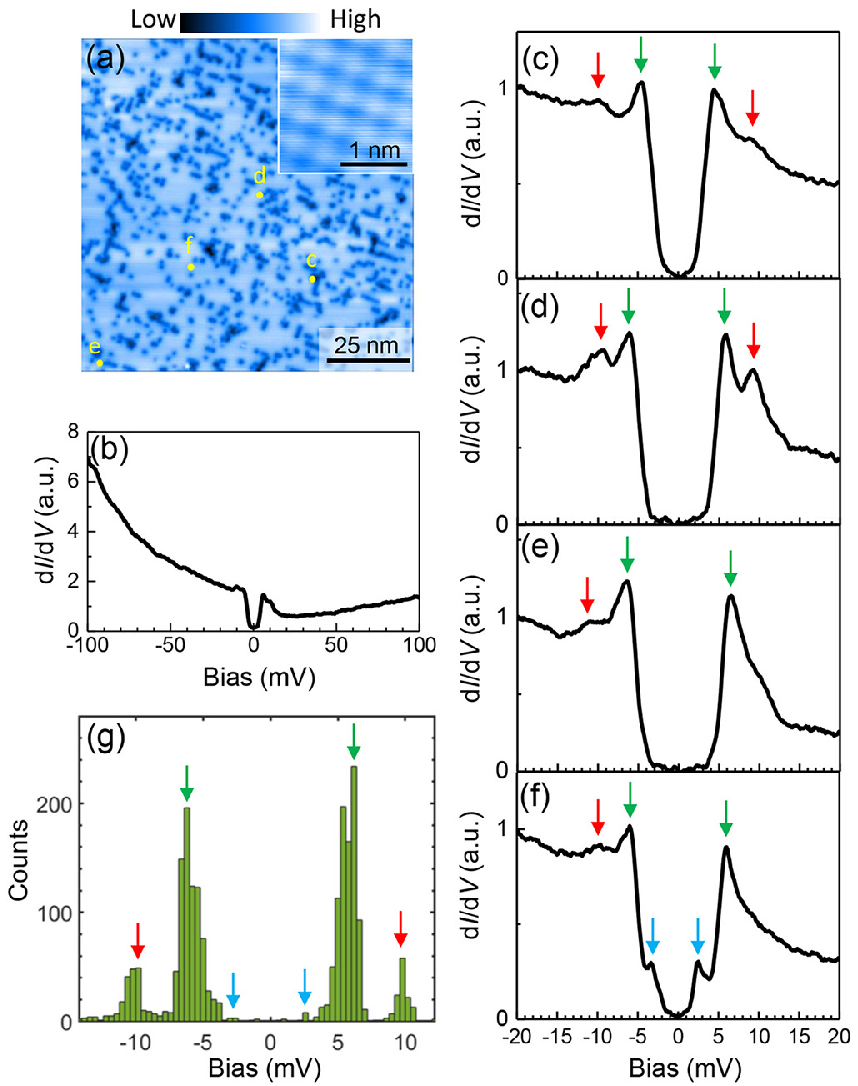}
\caption{(a) Topographic image of another kind of terminated surface ($V_\mathrm{set} = 1$ V, $I_\mathrm{set} = 20$ pA). Hollows in this topography have a lower density and a smaller averaged size when compared with ones in Fig.~\ref{fig2}(a). The inset shows the atomically resolved topography measured in the flat area far away from hollows ($V_\mathrm{set}$ = 100 mV, $I_\mathrm{set}$ = 200 pA). (b) A typical tunneling spectrum measured to high energy ($V_\mathrm{set}$ = 100 mV, $I_\mathrm{set}$ = 200 pA). (c-f) Tunneling spectra measured at marked positions in (a) ($V_\mathrm{set}$ = 30 mV, $I_\mathrm{set}$ = 200 pA). The characteristic peaks are marked by arrows. (g) The statistics of the peak energies derived from 900 spectra which are measured at points with a matrix of 30$\times$ 30 uniformly distributed in the area of (a) ($V_\mathrm{set}$ = 50 mV, $I_\mathrm{set}$ = 500 pA). All measurements are carried out at 1.7 K.
} \label{fig7}
\end{figure}

\begin{figure}[htbp]
\centering
\includegraphics[width=8cm]{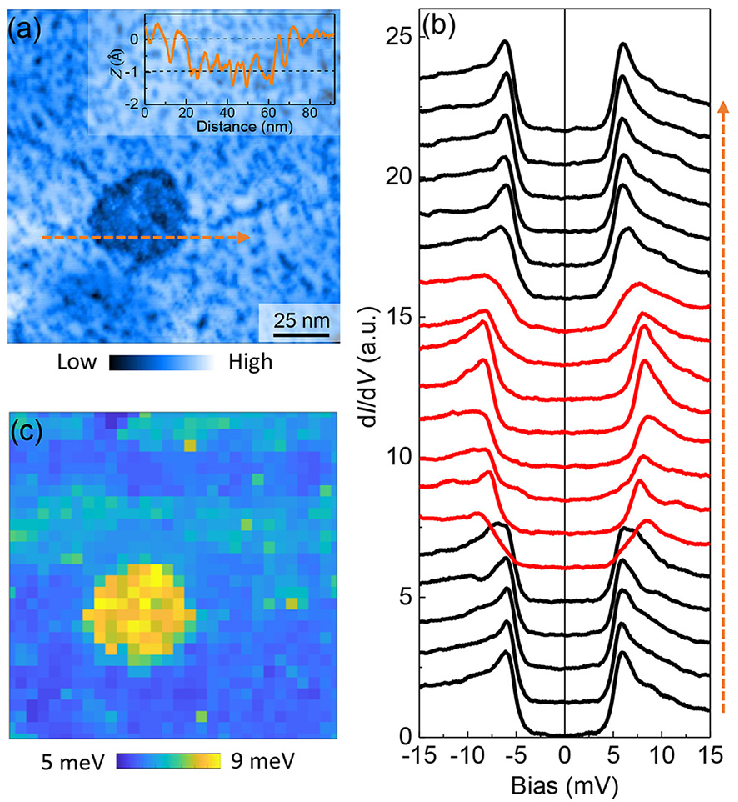}
\caption{(a) Topographic image with an exposed low-lying area of the under layer ($V_\mathrm{set}$ =50 mV, $I_\mathrm{set}$ = 100 pA). The inset shows the line profile of the surface height along the arrowed line, and the exposed low-lying layer is about 1 \AA\ by average lower than the top surface layer. (b) A set of tunneling spectra measured along the dashed line in (a) ($V_\mathrm{set}$ = 50 mV, $I_\mathrm{set}$ = 200 pA). (c) Color mapping of the coherence peak energy in positive bias side based on 900 tunneling spectra measured at points with a matrix of 30$\times$ 30 uniformly distributed on the area of (a) ($V_\mathrm{set}$ = 50 mV, $I_\mathrm{set}$ = 200 pA). All measurements are carried out at 1.7 K.
} \label{fig8}
\end{figure}

In the K12442 sample, we also observe other areas with different topographic features, and one example is shown in Fig.~\ref{fig7}(a). This kind of surface is rarely observed with a possibility of once in 8 times cleavage. On this surface the hollows are much smaller than the ones shown in Fig.~\ref{fig2}(a). However, from the atomically resolved surface shown in the inset in Fig.~\ref{fig7}(a), the lattice constant is also about 5.3 \AA\, which is close to the value obtained in Fig.~\ref{fig2}(a). Figure~\ref{fig7}(b) shows a tunneling spectrum measured in a wide bias range. The spectrum has the similar behavior comparing with the one shown in Fig.~\ref{fig2}(b): both of them show the particle hole asymmetry. The quantitative difference of them is the considerable bias-voltage dependence of d$I/$d$V$ in the negative bias side of the spectrum shown in Fig.~\ref{fig7}(b). This suggests slightly different band structures in these two kinds of surfaces. In the area of Fig.~\ref{fig7}(a), we can detect spectra with different gap energies, and four examples are shown in Figs.~\ref{fig7}(c)-\ref{fig7}(f). In the spectrum shown in Fig.~\ref{fig7}(f), we can also see low-energy peaks at about $\pm2.5$ meV which may be attributed to the impurity induced bound states. Besides, we can see that coherence peaks in spectra locate in a wider energy range compared to the ones shown in Fig.~\ref{fig2}(d). In addition, hump features can be observed at energy near $\pm10$ meV on some spectra. The humps on tunneling spectra may be the feature of a larger superconducting gap, or may be the bosonic mode which has been observed in many FeSCs \cite{Bosonicmode1,Bosonicmode2,LiFeAsMode}. Figure~\ref{fig7}(h) shows the energy statistic results of all the peak features based on 900 spectra measured in the area of Fig.~\ref{fig7}(a). One can see that the coherence peak feature are in the range of $\pm$(4-8) meV with possible local maxima at about $\pm4.4$, $\pm5.4$, and $\pm6.2$ meV. This suggests the multi-gap feature in the superconductor. The maximum probability of the hump feature appears at about $\pm9.8$ meV.

Figure~\ref{fig8}(a) shows a typical topography where a low-lying layer can be observed. The size of such exposed low-lying area is dozens of nanometers, and the average height of this area is about 1 \AA\, lower than the top surface. Then we measure the tunneling spectra across the exposed low-lying area and show them in Fig.~\ref{fig8}(b). One can see that the fully gapped feature can be observed on all the spectra, but energy values of the coherence peaks shift from about $\pm6$ meV on the top layer to about $\pm8$ meV in the exposed low-lying area. The feature is more clear in the coherence peak energy mapping shown in Fig.~\ref{fig8}(c) with much larger coherence peak energy in the exposed low-lying area.

\section{Discussions}\label{discussion}

On the surfaces of $\rm KCa_2Fe_4As_4F_2$ single crystals, we observe the $\sqrt{2}\times\sqrt{2}$ reconstructed layer of atoms. This is consistent with the fact that half of potassium atoms or calcium atoms of a layer should stay and reconstruct on each cleaved surface. This can lead to electrically unpolar surfaces. It should be noted that there are lots of hollows with the size of sub- or several nanometers randomly distributed on the surface. Such topography with hollows is very common in FeSCs with exposed surface composed by alkali metal atoms \cite{NaFeCoAs,LiFeAs,RbFe2As2}. The hollow population and distribution may be different in the surface terminated by alkaline-earth metal atoms. In this point of view, and considering the observed features, the commonly obtained top surface may be reconstructed by K atoms, while the rarely achieved surface may be reconstructed by the Ca atoms in K12442 samples. The hollows are either K or Ca vacancies in the two cases, which may be due to easy losing of these atoms in the cleaving procedure. The missing K/Ca atoms on the surface will adjust the band structure slightly, which is supported by the different features beyond the gap on spectra shown in Figs.~\ref{fig2}(b) and \ref{fig7}(b).

We observe obvious fully gapped feature in most tunneling spectra measured in K12442 single crystals, which indicates the absence of nodes and is consistent with other measurements \cite{ARPES,HeatTran,optical}. Most of the coherence peaks locate in the energy range from 4 to 8 meV, and these gap values seem to be comparable to those reported previously \cite{ARPES,optical,Hc1}. Since the dominant contribution of the FT-QPI is consistent with the intrapocket scattering of the $\alpha$ pocket, the obtained superconducting gap values on the easily achieved surface are likely to be assigned to this hole pocket. Therefor the superfluid may be mainly contributed by the hole-like Fermi surfaces near $\Gamma$ point. The situation is similar to that in CaKFe$_4$As$_4$: several hole and electron pockets are observed near $\Gamma$ and M points by the ARPES measurement \cite{ARPES1144}, but only the scattering between two hole pockets is observed in FT-QPI patterns from the STM measurement \cite{STM1144}. The hump feature at about $\pm9.8$ meV observed in K12442 on other type surface may be the larger superconducting gap as reported previously \cite{uSRK,Hc1} or the bosonic mode corresponding to some gap(s). If it is the bosonic mode, the mode energy is about 2-6 meV according to the substraction of the superconducting gap energy. However, the measured bosonic mode value is as large as 16 meV from neutron scattering experiments \cite{NSRK}, thus this feature most likely reflects an energy gap. Although we have observed some peaks on the spectra near $\pm$2.2 mV, due to the existence of some very strong impurity bound states appearing near this energy, we cannot assure that these features reflect a small superconducting gap\cite{ARPES,uSRK,Hc1}. From our experimental data, the confirmable superconducting gap ranges from 4 to 8 meV and the dominant contribution of superfluid may be from the bands with the gaps of about 4.4 to 5.4 meV, most probably the hole derivative $\alpha$ band near $\Gamma$. However, the determined Fermi energy of the $\alpha$ band is only about $24\pm6$ meV, this indicates that the basic requirement of the Bardeen-Cooper-Schrieffer (BCS) theory in the weak coupling limit, namely $E_\mathrm{F}\gg\Delta$ cannot be satisfied in K12442. This strongly suggests that the superconducting physics in K12442 material should possess by itself an unconventional feature.

In FeSCs, the scenario of $s^\pm$ pairing is based on the nesting between the hole and electron pockets with similar sizes in the weak coupling scenario. This pairing manner was firstly inferred from the QPI measurements on Fe(Se,Te) by comparing the difference when the applied magnetic field is zero and finite\cite{HanaguriScience}. This was later further strengthened by the impurity effect\cite{YangHNC2013} and phase resolved QPI analysis \cite{ChenMYPRBFeTeSe2019}. However, this pairing mechanism is challenged in K12442 because the electron pocket at M point is too small \cite{ARPES}, so the nesting condition cannot be satisfied. In our measurements, we cannot even observe the scattering pattern between the hole pockets and tiny electron pockets in the FT-QPI result. There are two possible reasons for this. The first one is the tunneling matrix element effect due to which we cannot detect the electron $\delta$ pocket. In fact we can only detect the intrapocket scattering of $\alpha$ pocket, and we cannot even detect the scattering related to $\beta$ and $\gamma$ pockets. Hence, it is not strange that we have not detected the scattering based on the $\delta$ pocket. The other possible reason is the low density of states for the $\delta$ pocket. However in any case, very different sizes of the hole and electron pockets challenge the nesting picture of $s^\pm$-pairing in the weak coupling scenario. From tunneling spectra obtained in this work, if the impurity is non-magnetic in nature as we argued, the impurity bound states at about $\pm2.2$ meV may suggest the sign change of superconducting gaps in this superconductor. In any case, it remains unclear but interesting to know what role is played by the shallow electron pocket and even the incipient hole bands \cite{ARPES} near the M point, and whether they help to form the ``incipient'' $s^\pm$ pairing \cite{incipient}.

\section{Conclusion}

In conclusion, by using scanning tunneling microscope, we have investigated the superconducting gaps and pairing mechanism of KCa$_2$Fe$_4$As$_4$F$_2$ single crystals. Most spectra exhibit a full gap feature with the gap value from 4 to 8 meV. On few spectra, some peaks can be observed at about $\pm2.2$ meV, which can be attributed to the impurity induced bound states. We have not seen the characteristic scattering pattern between hole to electron pockets based on the QPI data and related analysis, however the FT-QPI pattern can be well described by the intrapocket scattering of $\alpha$ pocket near $\Gamma$ point. The dispersion derived from the FT-QPI indicates a small Fermi energy of about 24 meV for the band forming the $\alpha$ pocket. This indicates a strong deviation from the basic requirement of the weak coupling BCS theory. Our results shed new light in helping to clarify the superconducting mechanism in iron based superconductors.

\begin{acknowledgments}
We appreciate useful discussions with A. V. Balatsky and Z. Y. Wang. This work was supported by National Key R\&D Program of China (Grants No. 2016YFA0300401, No. 2018YFA0704200, No. 2017YFA0303100, and No. 2017YFA0302900), National Natural Science Foundation of China (Grants No. 12061131001, No. 11974171, No. 11822411, No. 11961160699, No. 11674406, and No. 11674372), and the Strategic Priority Research Program (B) of Chinese Academy of Sciences (Grants No. XDB25000000, and No. XDB33000000). H. L. is grateful for the support from Beijing Natural Science Foundation (Grant No. JQ19002) and the Youth Innovation Promotion Association of CAS (Grant No. 2016004).
\end{acknowledgments}

$^*$ huanyang@nju.edu.cn

$^\dag$ hhwen@nju.edu.cn

\end{document}